\documentstyle[twoside,amssymb,12pt]{article}

\newcommand{\nc}{\newcommand}
\nc{\la}{\lambda} \nc{\alf}{\alpha}
\nc{\tht}{\theta}  \nc{\be}{\beta}  \nc{\eps}{\epsilon}
\nc{\ga}{\gamma}  \nc{\D}{\Delta}  \nc{\G}{\Gamma}  \nc{\vphi}{\varphi}
\nc{\de}{\delta} \nc{\si}{\sigma}  \nc{\ka}{\kappa}   \nc{\Si}{\Sigma}
\nc{\om}{\omega}  \nc{\Om}{\Omega}  \nc{\z}{\zeta}
\nc{\qq}{\quad\quad}   \nc{\nf}{\infty}   \nc{\pt}{\partial}
\nc{\dl}{\mathop{\smash{\cal L}}}  \nc{\black}{\rule{1mm}{3mm}}
\nc{\beq}{\begin{equation}}     \nc{\eeq}{\end{equation}}
\nc{\beqa}{\begin{eqnarray}}  \nc{\dst}{\displaystyle}  \nc{\sst}{\scriptstyle}
\nc{\eeqa}{\end{eqnarray}} \nc{\nnb}{\nonumber}
\nc{\bs}{\backslash}        \nc{\mb}{\mathbb}
\nc{\sn}{{\rm sn}\,} \nc{\cn}{{\rm cn}\,}     \nc{\dn}{{\rm dn}\,}
\nc{\ti}{\tilde}         \nc{\wti}{\widetilde}  \nc{\wh}{\widehat}
\nc{\ol}{\overline}      \nc{\ul}{\underline}
\nc{\modsim}{\mathop{\smash{\sim}}}     

\newcounter{muni}
\newenvironment{remunerate}{\begin{list}{{\rm \arabic{muni}.}}
{\usecounter{muni}
\setlength{\leftmargin}{0pt}\setlength{\itemindent}{38pt}}}{\end{list}}

\nc{\brm}{\begin{remunerate}}   \nc{\erm}{\end{remunerate}}

\newtheorem{nth}{Proposition}  
\nc{\simas}{\mathop{\smash{\sim}}}
\nc{\stg}{\mathop{\smash{*}}}   \nc{\LIM}{\mathop{\smash{\lim}}}
\nc{\st}{\mathop{\smash{\delta}}} \nc{\SUP}{\mathop{\smash{\rm sup}}}
\nc{\barr}{\begin{array}}   \nc{\earr}{\end{array}}   \nc{\dg}{\dagger}
\nc{\mtvb}{\mathversion{bold}}   \nc{\mtvn}{\mathversion{normal}}

\begin{document}
\begin{titlepage}

\vspace{5mm}
{\noindent \bf International Mediterranean Congress of Mathematics, \\ 
Almeria, 6-10 june 2005}

\vskip 2.0truecm 
\centerline{\Large\bf From asymptotics to spectral measures: }

\vspace{5mm}
\centerline{\Large\bf determinate versus indeterminate }

\vspace{5mm}
\centerline{\Large\bf moment problems}

\vspace{10mm}
\centerline{\Large\bf Galliano VALENT${}^{\;\dagger\; *}$}

\vskip 1.0truecm
\centerline{${}^{\dagger}$ \it Laboratoire de Physique Th\'eorique et des
Hautes Energies}
\centerline{\it CNRS, Unit\'e associ\'ee URA 280}
\centerline{\it 2 Place Jussieu, F-75251 Paris Cedex 05, France}
\nopagebreak

\vskip 0.5truecm
\centerline{${}^*$ \it D\'epartement de Math\'ematiques}
\centerline{\it UFR Sciences-Luminy}
\centerline{\it Case 901 163 Avenue de Luminy}
\centerline{\it 13258 Marseille Cedex 9, France}
\nopagebreak

\vskip 0.5truecm

\begin{abstract}
\noindent In the field of orthogonal polynomials theory, the classical Markov 
theorem shows that for determinate moment problems the spectral measure is under 
control of the polynomials asymptotics.

The situation is completely different for indeterminate moment problems, in which 
case the interesting spectral measures are to be constructed using Nevanlinna theory. 
Nevertheless it is interesting to observe that some spectral measures can still 
be obtained from weaker forms of Markov theorem.

The exposition will be illustrated by orthogonal polynomials related to 
elliptic functions: in the determinate case by examples due to Stieltjes and some of 
their generalizations and in the indeterminate case by more recent examples.
\end{abstract}

\end{titlepage}

\section{Background material}
Let us consider the three terms recurrence \footnote{We stick, as far as possible, to 
Akhiezer's notations in \cite{Ak}.}
\beq\label{b1}
xP_n=b_{n-1}\,P_{n-1}+a_nP_n+b_n\,P_{n+1},\qq n\geq 1.\\[4mm]
\eeq
We will denote by $P_n$ and $Q_n$ two linearly independent solutions of this 
recurrence with initial conditions
\beq\label{b2}
P_0(x)=1,\qq P_1(x)=\frac{x-a_0}{b_0},\qq\qq Q_0(x)=0,\qq Q_1(x)=\frac 1{b_0}.\eeq
The corresponding Jacobi matrix is 
\beq\label{Jacobi}
\left(\barr{cccccc}
a_0 & b_0 & 0 & 0 & 0 & \cdots\\[4mm]
b_0 & a_1 & b_1 & 0 & 0 & \cdots\\[4mm]
0 & b_1 & a_2 & b_2 & 0 & \cdots\\[4mm]
\cdots & \cdots & \cdots & \cdots & \cdots & \cdots\\[4mm]
\cdots & \cdots & \cdots & \cdots & \cdots & \cdots \earr\right)
\eeq
If the $b_{n}>0$ and $a_{n}\in{\mb R}$ the $P_n$ (resp. the $Q_n$) will be 
orthogonal with respect to a positive probabilistic measure $\psi$ (resp $\psi^{(1)}$)
\beq\label{b3}
\int \,P_m(x)\,P_n(x)\,d\psi(x)=\de_{mn},\qq\quad 
\int \,Q_m(x)\,Q_n(x)\,d\psi^{(1)}(x)=\de_{mn},\eeq
with the moments
\beq\label{b4}
s_n=\int x^n\,d\psi(x),\qq n\geq 0,\qq s_0=1\eeq
If ${\rm supp}\,\psi\subset [0,+\nf[$ we have a Stieltjes moment 
problem while if ${\rm supp}\,\psi\subset ]-\nf,+\nf[$ we have a Hamburger moment 
problem. These moment problems may be determinate (det S or det H) if the measure 
is unique or indeterminate (indet S or indet H) if it is not unique.

For further use, we will introduce new polynomials $F_n(x)$ by
\beq\label{mt1}
P_n(x)=\frac{(-1)^n}{\sqrt{\pi_n}}\,F_n(x),\qq n\geq 0,\qq
\barr{l}
a_n=\la_n+\mu_n,\qq\qq b_n=\sqrt{\la_n\mu_{n+1}},\quad n\geq 0,\\[4mm]\dst 
\pi_0=1,\qq\pi_n=\frac{\la_0\la_1\dots\la_{n-1}}{\mu_1\mu_2\cdots\mu_n},\quad n\geq 1,
\earr\eeq
 From (\ref{b1}) we deduce
\beq\label{mt3}\barr{l}
-xF_n=\mu_{n+1}F_{n+1}+(\la_n+\mu_n)F_n+\la_{n-1}F_{n-1},\\[4mm]
 F_{-1}(x)=0,\quad F_0(x)=1,\earr
\eeq
Similarly, defining
\beq\label{mt4}
Q_n(x)=\frac{(-1)^{n-1}}{\mu_1\sqrt{\pi_n}} \,F^{(1)}_{n-1}(x),
\eeq
one can check from (\ref{b1}) that the $F^{(1)}_n(x)$ are a solution of the 
recurrence (\ref{mt3}) with the substitution 
$\,(\la_n,\,\mu_n)\ \to\ (\la_{n+1},\mu_{n+1}),$ i. e. the associated 
polynomials of order one.

Notice the useful relations and notations, valid for $\mu_0=0$, easily derived 
by induction
\beq\label{mt5}
P_n(0)=(-1)^n\sqrt{\pi_n},\qq\qq Q_n(0)=(-1)^n\frac{\sqrt{\pi_n}}{\alf_n},\qq\qq 
\frac 1{\alf_n}=-\sum_{k=1}^n\frac 1{\mu_k\pi_k}.\eeq

\[  \]
\centerline{\Large\bf The determinate case}

\section{Markov theorem}
In the determinate case (det H hence det S), given $(a_n,b_n)$ the basic tool 
to compute the spectral measure is Markov theorem. In the classical textbooks 
\cite[\S 3.5]{Sz},\cite[p. 89]{Ch1} it is proved under the restrictive assumption 
that the measure support is bounded (which implies that the moment problem is 
determinate). More recently it was proved under the sole hypothesis of determinacy 
of the moment problem \cite{Be},\cite{VAs}. It can be stated as :
\begin{nth}
For a determinate moment problem the Stieltjes transform of the (unique) 
orthogonality measure is given by
\beq\label{markov}
\lim_{n\to\nf} \ \frac{Q_{n}(x)}{P_n(x)}=
-\lim_{n\to\nf} \ \frac{F^{(1)}_{n-1}(x)}{\mu_1\, F_n(x)}=
\int\frac{d\psi(t)}{x-t},\qq\quad x\in V={\mb C}\backslash{\mb R},\eeq
where the convergence is uniform in compact subsets of $V.$ 
\end{nth}

Let us mention the connection with finite continued fractions. One has 
\beq\label{cf1}
\frac{Q_{n}(x)}{P_n(x)}= 1/x-a_0-b_0^2/x-a_1-b_1^2/x-\cdots
-b_{n-2}^2/x-a_{n-1},
\eeq
which can be written, using the Christoffel numbers $\la_{k,n}$ according to
\beq\label{cf2}
\frac{Q_{n}(x)}{P_n(x)}= \sum_{k=1}^n\frac{\la_{k,n}}{x-x_{k,n}}=
\int\frac{d\psi_n(t)}{x-t}.\eeq
As shown in \cite{Be}, when the moment problem is determinate, the measure 
$\psi_n$ converges weakly to $\psi.$ The limiting continued fraction does give the 
Stieltjes transform of the spectral measure
\beq\label{cf3}
\int\frac{d\psi(t)}{x-t}=1/x-a_0-b_0^2/x-a_1-b_1^2/x-\cdots,\qq 
x\in {\mb C}\backslash{\mb R}.\eeq
The continued fraction encodes not only the coefficients appearing in the recurrence 
relation of the polynomials, but also the moments in the asymptotic series 
\beq\label{cf5}
\int\frac{d\psi(t)}{x+t}\asymp \sum_{n\geq 0}(-1)^n\frac{s_n}{x^{n+1}},
\eeq
valid uniformly in $\de\leq{\rm arg}\, x\leq \pi-\de$ provided that $0<\de<\pi/2,$ 
as shown in \cite[p. 95]{Ak}.

Since we want to discuss some work of Stieltjes, let us mention that he often 
substitutes 
\[ x\to -x^2,\qq a_n=\la_n+\mu_n,\qq b_n^2=\la_n\,\mu_{n+1},\]
 and writes
\beq\label{cf4}
\int\frac{d\psi(t)}{x^2+t}=
1/x^2+\la_0+\mu_0-\la_0\mu_1/x^2+\la_1+\mu_1-\la_1\mu_2/x^2+\cdots,\qq 
x\in {\mb C}\backslash{\mb R}.\eeq
Let us make the further assumption that $\mu_0=0.$ Considering
\[\int\frac{x}{x^2+t}d\psi(t)=
x(1/x^2+\la_0-\la_0\mu_1/x^2+\la_1+\mu_1-\la_1\mu_2/x^2+\cdots),\]
and upon iteration of the identity \cite[p. 404]{St2}
\[x^2+\la_0/1+\mu_1/D=x^2+\la_0-\la_0\mu_1/\mu_1+D,\] 
we get first
\[\int\frac{x}{x^2+t}d\psi(t)=1/x^2+\la_0/1+\mu_1/x^2+\la_1/1+\mu_2/x^2+\cdots,\]
easily transformed into
\beq\label{sform}
\int\frac{x}{x^2+t}d\psi(t)=1/x+\la_0/x+\mu_1/x+\la_1/x+\mu_2/x+\cdots\eeq
So, under the assumption that $\mu_0=0,$ we have transformed the initial 
J-continued fraction (\ref{cf3}) into an S-continued fraction (\ref{sform}), 
following the derivation due to Stieltjes in \cite{St2}.

\section{Stieltjes continued fractions with elliptic functions}
Stieltjes gave four continued fractions involving the Jacobi elliptic functions 
usually denoted as  
${\rm sn}\,(u,k^2),\,{\rm cn}\,(u,k^2)$ and $\,{\rm dn}\,(u,k^2),$ 
with parameter $0<k^2<1.$ Let us record two of them
\beq\label{st1}\barr{l}\dst
\int_0^{\nf}\,{\rm dn}\, u \,e^{-xu}\,du= 1/x+k^2/x+2^2/x+3^2k^2/x+4^2/x+\cdots,\\[4mm]
\dst\int_0^{\nf}\,{\rm cn}\, u \,e^{-xu}\,du= 1/x+1^2/x+2^2k^2/x+3^2/x+4^2k^2/x+\cdots
\earr\eeq
for ${Re}\,x>0.$ These relations are also quoted in Wall's book \cite[\S 94]{Wa}.

On these relations we recognize S-continued fractions, the first one corresponding 
to the polynomials with recurrence coefficients $\la_n=k^2(2n+1)^2\ \ \mu_n=4n^2,$ 
and the second one to $\la_n=(2n+1)^2\ \ \mu_n=4k^2n^2.$ 

Notice that using the transformation theory of elliptic functions, namely the 
relation $\,{\rm dn}\,(u;k)={\rm cn}\,(ku;1/k),$ one can deduce, by elementary 
algebra, the second continued fraction from the first one.

The first proof of (\ref{st1}), published by Stieltjes in 1889 in \cite{St1}, used 
intensively the addition relations for the elliptic Jacobi functions and was 
quite lengthy (it may be found in Wall's book). But in 1891, in a  letter to 
Hermite (published only in 1905 \cite[p. 208]{BB}), he found an elegant shorter proof 
which we shall report \footnote{Exactly the same proofs appear in \cite{Rog}, without 
any reference to Stieltjes, but some years later, in 1907.}

The starting point is to define
\beq\label{fc1}
C_n=\int_0^{\nf}\,{\rm cn}\,u\,({\rm sn}\,u)^n\,e^{-xu}\,du,\qq
D_n=\int_0^{\nf}\,{\rm dn}\,u\,({\rm sn}\,u)^n\,e^{-xu}\,du,\qq n\in{\mb N}.\eeq
For ${\rm Re}\,x>0$ an integration by parts gives 
\beq\label{fc2}\barr{lll}
xC_0=1-D_1,    &\qq xC_n=nD_{n-1}-(n+1)D_{n+1}, &\qq n\geq 1,\\[4mm]
xD_0=1-k^2C_1, &\qq xD_n=nC_{n-1}-k^2(n+1)C_{n+1}, &\qq n\geq 1.\earr\eeq
So if we define 
\beq\label{fc3}
p_0=C_0,  \qq p_n=\frac{C_n}{nD_{n-1}},\qq q_0=D_0,\qq q_n=\frac{D_n}{nC_{n-1}},\quad 
n\geq 1,\eeq
we get the non-linear recurrences
\beq\label{fc4}
p_n=\frac 1{x+(n+1)^2\,q_{n+1}},\qq\qq q_n=\frac 1{x+k^2(n+1)^2\,p_{n+1}},\qq n\geq 0.
\eeq
Iterating these relations starting from $p_0$ and $q_0$ gives relations (\ref{st1}).

The continued fractions given by Stieltjes are quite impressive, since from them we  
can get easily the moments and the orthogonality measure, as we will explain now. 

Let us start from the Taylor series
\beq\label{mom1}\left\{\barr{l}\dst 
{\rm dn}\,u=\sum_{n\geq 0}(-1)^n\,\frac{s_n}{(2n)!}\,u^{2n},\\[4mm]
s_0=1,\qq s_1=k^2,\qq s_2=k^2(4+k^2),\qq s_3=k^2(16+44k^2+k^4),\qq\cdots\earr\right.
\eeq
which, inserted in (\ref{st1}), induces the asymptotic series
\beq\label{mom2}
\int \frac x{x^2+t}d\psi(t)= \int_0^{\nf}\,{\rm dn}\, u \,e^{-xu}\,du \asymp 
\sum_{n\geq 0}(-1)^n\frac{s_n}{x^{2n+1}},\eeq
from which we conclude that the coefficients $s_n$ are indeed the moments of $\psi.$ 
Their asymptotics follows easily from the generating function (\ref{mom1}) and Darboux 
theorem: 
\beq\label{mom3}\dst 
s_n\ \modsim_{n\to\nf}\ 2\frac{(2n)!}{(K')^{2n+1}},\eeq
showing explicitly that the series (\ref{mom2}) is indeed asymptotic.

Let us start from the Fourier series 
\beq\label{om1}
{\rm dn}\,u=\psi_0+\sum_{n\geq 1}\psi_n\,\cos\left(n\frac{\pi u}{K}\right),
\eeq 
with the coefficients
\beq\label{om3}
\psi_0=\frac{\pi}{2K},\qq\quad\psi_n=\frac{2\pi}{K}\,\frac{q^n}{1+q^{2n}},
\quad n\geq 1,\qq q=e^{-\pi K'/K}.\eeq
Inserting this relation into the first continued fraction (\ref{st1}) gives
\beq\label{om4}
\int\,\frac{x}{x^2+t}\,d\psi(t)=\frac{\psi_0}{x}
+\sum_{n\geq 1}\psi_n\ \frac x{x^2+(n\pi/K)^2},\eeq
showing that the spectral measure is discrete 
\beq\label{om5}
\psi=\sum_{n\geq 0}\ \psi_n\,\eps_{(n\pi/K)^2},\eeq
where $\eps_s$ is the discrete measure with unit jump. 
Similar results can be obtained for the first continued fraction in (\ref{st1}).

These deep and elegant results of Stieltjes are quite frustrating since they 
apparently don't bear any relation with asymptotics. So how should we proceed 
to derive Stieltjes results using Markov theorem?

\section{Stieltjes continued fractions from Markov theorem}\label{mkv}
Let us consider the continued fraction with $\la_n=k^2(2n+1)^2$ and $\mu_n=4n^2.$ 
We need the asymptotics of the polynomials $F_n$ and of their 
associates of order one $F^{(1)}_n.$ So we need {\em two} generating functions. 
Carlitz \cite{Ca} has obtained a first one
\beq\label{gfF}
F(x;w)\equiv \sum_{n\geq 0}\,\frac{n!}{(1/2)_n}\,F_n(x)\,w^n=
\frac{\cos(\sqrt{x}\tht(w))}{\sqrt{1-k^2w}},\qq \tht(w)=
\int_0^w\,\frac{du}{2\sqrt{u(1-u)(1-k^2u)}}.\eeq
Notice, en passant, that $G(x;w)=\sqrt{1-k^2 w}\,F(x;w)$ is a solution of Heun's 
differential equation \cite{Ron}
\beq\label{Hn}
\frac{d^2G}{dw^2}+\left(\frac{1/2}{w}-\frac{1/2}{1-w}-\frac{k^2/2}{1-k^2 w}\right)
\frac{dG}{dw}+\frac x4\,G=0.\eeq
Using theorem (8.4) in \cite{Sz} (see \cite{Va1} for the details) one deduces 
the asymptotics
\beq\label{asF}
F_n(x)\sim -\frac 1{2k'^2\,n}\,\frac{\pi_n}{(k^2)^n}\,\sqrt{x}\sin(\sqrt{x}K),\qq 
x\in{\mb C}\backslash{\mb R}.\eeq
The generating function needed for the associated polynomials $F_n^{(1)}$ was given 
in \cite{Va1} (set $c=1$ and $\mu=0$ in the relation (2.15) of this reference):
\beq\label{gfF1}
\sum_{n\geq 0}\frac{(2)_n}{(3/2)_n}w^{n+1}\,\frac{F_n^{(1)}(x)}{\mu_1}=
\frac{N(w)}{2\sqrt{1-k^2 w}},
\eeq
with
\beq\label{N}
N(w)=\int_0^{\tht(w)}\frac{\sin(\sqrt{x}(\tht(w)-u))}{\sqrt{x}} \,{\rm dn}\,u\,du.
\eeq
Darboux theorem gives 
\beq\label{asF1}
\frac{F_n^{(1)}(x)}{\mu_1}\sim -\frac 1{2k'^2\,n}\,\frac{\pi_n}{(k^2)^n}\,
\int_0^K\,\cos(\sqrt{x}(K-u))\,{\rm dn}\,u\,du, \qq 
x\in{\mb C}\backslash{\mb R}.\eeq

We can now use Markov theorem to obtain
\[\dst\int\frac{d\psi(t)}{x-t}=
\frac{\int_0^K\, {\rm dn}\,u\cos(\sqrt{x}(K-u))\,du}{\sqrt{x}\sin(\sqrt{x}K)},
\qq x\in{\mb C}\backslash{\mb R}.\]
Let us reduce this result to its Stieltjes form. We first substitute 
$x\Rightarrow -x^2$ which gives
\[ \dst\int\frac{x}{x^2+t}d\psi(t)=
\frac 1{\sinh(xK)}\,\int_0^K\,{\rm dn}\,u\cosh(x(K-u))\,du.\]
The change of variables $v=2K-u$ allows to show
\[\int_0^K\,e^{-x(K-u)}\,{\rm dn}\,u\,du=\int_K^{2K}\,e^{x(K-v)}\,{\rm dn}\,v\,dv,\]
and this implies
\[\int_0^K\,{\rm dn}\,u\cosh(x(K-u))\,du=e^{xK} \int_0^{2K}\,e^{-xu}\,{\rm dn}\, u\,du.\]
It follows for the Stieltjes transform that
\[ \int\frac{x}{x^2+t}d\psi(t)=
\frac 1{1-e^{-2xK}}\,\int_0^{2K}\,e^{-xu}\,{\rm dn}\, u\,du =
\int_0^{\nf}\,{\rm dn}\,u\, e^{-xu}\,du,\qq {\rm Re}\,x>0.\]
The last equality follows from the $2K$-periodicity of ${\rm dn}\,u.$ So, quite 
satisfactorily, Markov theorem reproduces Stieltjes results, certainly not so 
elegantly, but with the possibility of some generalizations which would be quite 
difficult remaining in Stieltjes approach.

\section{Generalization of Stieltjes results}
Since Stieltjes results in the nineteenth century, only a few generalizations 
could be obtained. The first one is due to the Chudnowski \cite{CC}, who changed 
the elliptic function $f(u)={\rm dn}\,u$ into solutions of Lam\'e's equation
\[\frac{d^2 f}{du^2}+xf=n(n+1)k^2\,{\rm sn}^2\,u\,f,\qq n\in{\mb N},\]
but no explicit results were given on the spectral measure and, since $n$ is an 
integer, there is no limiting process which can lead back to Stieltjes continued 
fractions (\ref{st1}). 

Another generalization, involving a continuous parameter $c>0$, was obtained  
in \cite{Va1}. Working out an appropriate generating function and the polynomials 
asymptotics, Markov theorem \footnote{Use relations given page 756 in the previous 
reference, and algebraic steps as in section \ref{mkv}.} 
yields :

\begin{nth}
For the orthogonal polynomials with recurrence coefficients 
\beq\label{rec}
\la_n=k^2(2n+2c+1)^2,\qq\mu_n=4(n+c)^2(1-\de_{n0}),\qq n\geq 0,
\eeq
the Stieltjes transform of the orthogonality measure is given, for 
$\,x\in{\mb C}\backslash{\mb R}\,$ and $\ c>0$, by 
\beq\label{genst1}
\int\frac{x}{x^2+t}d\psi(t)=\frac{N(c;x)}{D(c;x)}=1/x+\la_0/x+\mu_1/x+\la_1/x+\mu_2/x+
\cdots,\eeq
with
\beq\label{genst2}
N(c;x)=\int_0^{2K}\, {\rm dn}\,u\ \frac{({\rm sn}\,u)^{2c}}{(2c)!}\,e^{-xu}\,du,\qq 
D(c;x)=
\int_0^{2K}\, {\rm cn}\,u\ \frac{({\rm sn}\,u)^{2c-1}}{(2c-1)!}\,e^{-xu}\,du,
\eeq
using the notation $\,(\alf)!=\G(\alf+1).$ 
\end{nth}

\noindent{\bf Remarks:}

\brm
\item The limit $c\to 0$ is tricky for $D.$ One has to use
\[\lim_{c\to 0}D(c;x)=\lim_{c\to 0}\,2e^{-xK}\int_0^K\,\sinh(x(K-u))\,
{\rm cn}\,u\ \frac{({\rm sn}\,u)^{2c-1}}{(2c-1)!}\,du=2e^{-xK}\sinh(xK),\]
and in that way Stieltjes result is recovered, but we see that for 
a generic value of $c$ it is no longer possible to transform 
this ratio of integrals into a single integral. 
\item Their spectral properties are now under investigation \cite{RSV}: it can be 
shown that the spectrum is discrete and that its asymptotic behaviour is 
independent of the parameter $c.$
\erm 

\[  \]
\centerline{\Large\bf The indeterminate case}

\section{The Nevanlinna parametrization}
According to the growth of the coefficients $(\la_n,\,\mu_n),$ with $\mu_0=0,$ we may 
have three different possibilities \cite{Ak}:
\brm
\item indet S iff $\,\sum_{n=1}^{\nf}\left(\pi_n+1/\mu_n\pi_n\right)<\nf.$ 
\item indet H (which implies indet S) iff 
$\,\sum_{n=1}^{\nf}\pi_n(\sum_{k=1}^n1/\mu_k\pi_k)^2<\nf.$
\item det S and indet H iff $\,\sum_{n=1}^{\nf}1/\mu_n\pi_n=\nf$ and  
$\,\sum_{n=1}^{\nf}\pi_n(\sum_{k=1}^n1/\mu_k\pi_k)^2<\nf.$
\erm

For an indeterminate moment problem (see a detailed account in \cite{BV}), one 
first defines the series
\beq\label{Nesuites}\barr{ll}\dst 
A_n(x)= x\sum_{k=0}^{n-1}\,Q_k(0)\,Q_k(x), &\qq \dst 
C_n(x)=1+x\sum_{k=0}^{n-1}\,P_k(0)\,Q_k(x),\\[4mm]\dst 
B_n(x)=-1+x\sum_{k=0}^{n-1}\,Q_k(0)\,P_k(x), &\qq\dst  
D_n(x)=x\sum_{k=0}^{n-1}\,P_k(0)\,P_k(x),\earr \eeq
constrained by
\[ A_n(x)D_n(x)-B_n(x)C_n(x)=1.\]
In the indet H case, these series, for $n\to\nf$, converge absolutely and uniformly 
\cite{Ak} on compact subsets of ${\mb C}$) to entire functions $A(x),\cdots,D(x).$

The Nevanlinna matrix ${\cal N}$ is then
\beq\label{Ne1}
{\cal N}(x)=\left(\barr{cc} A(x) & C(x)\\[4mm] B(x) & D(x)\earr\right),\qq
A(x)D(x)-B(x)C(x)=1,\qq \forall x\in{\mb C}.\eeq
It gives the Stieltjes transform of all the Nevanlinna-extremal 
(or N-extremal) measures 
\beq\label{Ne3}
\int\frac{d\psi_{\la}(t)}{x-t}=\frac{A(x)\la-C(x)}{B(x)\la-D(x)},\qq 
\la\in{\mb R}\cup\{\nf\}\sim S^1.\eeq
For these measures and only for these measures are the polynomials $P_n$ 
dense in $L^2({\mb R},d\psi_{\la}).$ 

Let us observe that the Stieltjes transform being meromorphic, the N-extremal 
measures are all discrete with
\beq\label{Ne4}
\psi_{\la}=\sum_{s\in Z_{\la}}\psi_{\la}(s)\,\eps_s,\qq \qq 
\psi_{\la}(s)=\frac 1{B'(s)D(s)-B(s)D'(s)}\eeq
where $Z_{\la}$ is the zero set of the entire function $B\la-D$ (or $B$ for $\la=\nf$).

The series 
\beq\label{Ne5}\dst 
\frac 1{\alf_n}=\frac{Q_n(0)}{P_n(0)}= -\sum_{k=1}^{n}\frac 1{\mu_k\pi_k},\qq 
\frac 1{\alf}=\lim_{n\to\nf}\,\frac 1{\alf_n}=-\sum_{k=1}^{\nf}\frac 1{\mu_k\pi_k},
\eeq 
is quite important since, as shown in \cite{Ch2}, \cite{BV} the positively supported 
measures are given by $\la\in[\alf,0].$ As we will see the border measures 
$\psi_0$ and $\psi_{\alf}$ play a prominent role. In terms of the self-adjoint 
extensions of the Jacobi matrix $\psi_0$ corresponds to Krein's extension \cite{Si} 
and $\psi_{\alf}$ corresponds to Friedrichs extension \cite{Pe}. 

Polynomials for which the Nevanlinna matrix and N-extremal measures are known, more or 
less explicitly, are not very numerous: they correspond to strong increase of the 
$(\la_n,\mu_n)$ for large $n$. This increase may be exponential, as  
for the $q^{-1}$-Hermite \cite{IM}, and in this case all the N-extremal 
measures are known explicitly! Many other references to related to q-polynomials are 
given in \cite{BV}.

When the $(\la_n,\mu_n)$ are some particular quartic polynomial \cite{BV} 
the Nevanlinna matrix and  the border N-extremal measures are explicitly 
known. More recently the Nevanlinna matrices for some cubic cases have been 
obtained \cite{GLV} but only the asymptotics of the N-extremal spectra 
could be obtained. An example of the ``exotic" case det S and indet H is 
available for the Al-Salam-Carlitz polynomials and is discussed in \cite{BV}.

Let us now turn to the determination of the Nevanlinna matrix from 
generating functions.

\section{Dual polynomials versus Nevanlinna matrix}
Using the relations given in (\ref{mt1}), (\ref{mt4}) and (\ref{mt5}) the Nevanlinna 
matrix can be written as
\beq\label{dp1}
\left\{\barr{ll}\dst 
A(x)=-\frac x{\mu_1}\sum_{n=1}^{\nf}\frac{F_{n-1}^{(1)}(x)}{\alf_{n}}, &\qq\dst  
B(x)=-1+x\sum_{n=1}^{\nf}\frac{F_n(x)}{\alf_n},\\[5mm]\dst 
C(x)=1-\frac x{\mu_1}\sum_{n=0}^{\nf}F_n^{(1)}(x),&\qq\dst 
D(x)=x\sum_{n=0}^{\nf}F_n(x).
\earr\right.\eeq
If we know the generating function 
$G(x;w)=\dst\sum_{n\geq 0}F_n(x)w^n,$ from Abel's lemma we deduce
\[D(x)=x\,\lim_{w\to 1-}\ G(x;w),\]
and similarly for the function $C$ related to the polynomials $F^{(1)}_n.$ 

The computation of $A$ and $B,$ as shown in \cite{Va2}, is related to the dual 
polynomials $\wti{F}_n$ defined in \cite{KMG} by the recurrence
\beq\label{dp2}\barr{l}
-x\wti{F}_n=\wti{\mu}_{n+1}\wti{F}_{n+1}+(\wti{\la}_n+\wti{\mu}_n)\wti{F}_n+
\wti{\la}_{n-1}\wti{F}_{n-1},\\[4mm]
 \wti{F}_{-1}(x)=0,\quad \wti{F}_0(x)=1,\earr
\eeq
with the coefficients \cite{KMG}
\beq\label{dprec}
\wti{\la}_n=\mu_{n+1},\quad n\geq 0,\qq\quad \wti{\mu}_n=\la_n,\quad n\geq 0,\qq
\wti{\pi}_n=\frac{\la_0}{\mu_{n+1}\pi_{n+1}}.
\eeq
Notice that for the initial coefficients $(\la_n,\,\mu_n)$ we have $\mu_0=0,$ but for 
the dual coefficients $\wti{\mu}_0=\la_0>0$ from positivity. 

Let us prove first:

\begin{nth}
Let us consider an indet S moment problem, with coefficients $(\la_n,\,\mu_n)$ such 
that $\mu_0=0.$ Let the $\wti{F}_n$ be the dual polynomials as defined previously. 
Then one has
\beq\label{dp3}
B(x)-\frac{D(x)}{\alf}=-1+\frac x{\wti{\mu}_0}\sum_{n\geq 0}\,\wti{F}_n(x).
\eeq
\end{nth} 

\noindent{\bf Proof:}

\noindent Let us start from the double series for $B$ given in (\ref{dp1}). Since the 
moment problem is indet S, the series $\dst -\frac 1{\alf_n}$ is absolutely 
convergent and the same is true for the series $\dst\sum_n\,F_n(x)$ for $x$ in any 
compact subset of ${\mb C}.$ We can interchange the order of the summations to get
\beq\label{dp4}
B(x)= -1-x\sum_{k\geq 1}\frac 1{\mu_k\pi_k}\,\sum_{n\geq k}F_n(x)=-1
-x\sum_{k\geq 1}\frac 1{\mu_k\pi_k}
\left(\sum_{n\geq 0}F_n(x)-\sum_{n=0}^{k-1}F_n(x)\right).
\eeq
The first piece is related to the function $D$ and the second one is simplified 
using the relation, proved by induction:
\beq\label{dp5}
\sum_{n=0}^{k-1}F_n(x)=\frac 1{\wti{\pi}_{k-1}}\,\wti{F}_{k-1}(x)=
\frac{\mu_k\pi_k}{\wti{\mu}_0}\,\wti{F}_{k-1}(x),
\eeq
and this concludes the proof. $\qq\Box$

Let us define the zero-related dual polynomials $\wh{F}_n$ as those polynomials with 
recurrence coefficients
\beq\label{dpco}
\wh{\la}_n=\wti{\la}_n=\mu_{n+1},\quad n\geq 0,\qq\quad 
\wh{\mu}_n=\wti{\mu}_n(1-\de_{n0})=\la_n(1-\de_{n0}).
\eeq
These new polynomials can be expressed in terms of the $\wti{F}_n$ and their associates 
of order one by
\[ \wh{F}_n(x)=\wti{F}_n(x)-\frac{\wti{\mu}_0}{\wti{\mu}_1}\wti{F}^{(1)}_{n-1}(x),\qq 
n\geq 0.\]

We are now in position to prove:

\begin{nth}
Let us consider an indet S moment problem, with coefficients $(\la_n,\,\mu_n)$ such 
that $\mu_0=0.$ Let the $\wh{F}_n$ be the zero-related dual polynomials as 
defined above. Then one has
\beq\label{dp6}
A(x)-\frac{C(x)}{\alf}=\frac 1{\wti{\mu}_0}\sum_{n\geq 0}\,\wh{F}_n(x).
\eeq
\end{nth} 

\noindent{\bf Proof:}

\noindent Let us start from the double series for $A$ given in (\ref{dp1}). By the  
same arguments as in the previous proposition, we can interchange the order 
of the summations to get
\beq\label{dp7}
A(x)= \frac x{\mu_1}\sum_{k\geq 1}\frac 1{\mu_k\pi_k}
\left(\sum_{n\geq 1}F^{(1)}_{n-1}(x)-\sum_{n=1}^{k-1}F^{(1)}_{n-1}(x)\right).
\eeq
The first piece is related to the function $C$ and the second one is simplified 
using the relation, proved by induction:
\beq\label{dp8}
-\frac x{\mu_1}\sum_{n=1}^{k-1}F^{(1)}_{n-1}(x)=
-1+\frac 1{\wti{\pi}_{k-1}}\wh{F}_{k-1}(x),
\eeq
and, taking into account $\mu_k\pi_k\wti{\pi}_{k-1}=\wti{\mu}_0,$ this 
concludes the proof. $\qq\Box$

To conclude this section, it seems interesting to modify slightly the 
Nevanlinna matrix ${\cal N}$ to the form
\beq\label{Nemod1}
\wti{\cal N}(x)=\left(\barr{cc}\wti{A}(x) & \wti{C}(x)\\[4mm]
\wti{B}(x) & \wti{D}(x)\earr\right),\qq 
\barr{ll}\dst 
\wti{A}=A-\frac{C}{\alf}, & \qq \wti{C}=C,\\[4mm]\dst 
\wti{B}=B-\frac D{\alf}, & \qq \wti{D}=D,\earr \qq \det\wti{\cal N}=1.
\eeq
Then the Stieltjes transform, defining $\mu=\alf\la/(\la-\alf)$, becomes
\beq\label{Nemod2}
\int\frac{d\psi_{\mu}(t)}{x-t}=
\frac{\wti{A}(x)\mu-\wti{C}(x)}{\wti{B}(x)\mu-\wti{D}(x)},\qq 
\mu\in{\mb R}\cup\{\nf\}\sim S^1.\eeq
The positively supported measures correspond now to $\mu \in{\mb R}^+\cup\{\nf\}$, 
and the border measures $(\psi_{\alf},\,\psi_0)$ become $(\psi_{\nf},\,\psi_0).$

\section{Markov-like theorems}
Despite Nevanlinna theory, which describes all the measures, the question 
of what survives from Markov theorem remains interesting . As we will see, the 
two border measures $\psi_{\alf}$ and $\psi_0$ 
are still given by Markov-like theorems. Indeed one has first:

\begin{nth}
For an indeterminate Stieltjes moment problem we have
\beq\label{mk1}
\lim_{n\to\nf}\,\frac{Q_n(x)}{P_n(x)}=
- \lim_{n\to\nf}\,\frac 1{\mu_1} \,\frac{F^{(1)}_{n-1}(x)}{F_n(x)}=
\int\frac{d\psi_{\alf}(t)}{x-t},\qq x\in V={\mb C}\backslash{\mb R}\eeq
where the convergence is uniform for $x$ in any compact subset of $V.$
\end{nth}

\noindent{\bf Proof:}

\noindent The proof given in \cite{Be} follows easily from two relations proved 
in \cite[p. 14]{Ak}, which may be written
\beq\label{mk2}
Q_n(x)=Q_n(0)C_n(x)-P_n(0)A_n(x),\qq\quad P_n(x)=Q_n(0)D_n(x)-P_n(0)B_n(x).
\eeq
We can replace $P_n(0)/Q_n(0)$ by $\alf_n$ (see relation (\ref{mt5})) so that
\beq\label{mk3}
\frac{Q_n(x)}{P_n(x)}=\frac{A_n(x)\,\alf_n-C_n(x)}{B_n(x)\,\alf_n-D_n(x)}.\eeq
For $n\to\nf$, since we are indet S, we have $\dst\lim_{n\to\nf}\,\alf_n=\alf$ and the 
series $A_n(x),\cdots,D_n(x)$ converge uniformly in compact subsets of 
$V$ to the entire functions $A(x),\cdots,D(x).$ It follows that
\beq\label{mk4}
\lim_{n\to\nf}\ \frac{Q_n(x)}{P_n(x)}=\frac{A(x)\,\alf-C(x)}{B(x)\,\alf-D(x)}.\eeq
The theorem follows from (\ref{Ne3}).$\qq\Box$

\vspace{2mm}
\noindent{\bf Remark:} If the moment problem is det S but indet H, then 
$\psi_{\alf}=\psi_0,$ is the {\em unique} measure supported by $[0,+\nf[$ 
(the previous theorem does still work in this case), while there are plenty 
of different measures supported by ${\mb R}$ and given by (\ref{Ne3}) for $\la\neq 0.$

Let us give another Markov-like theorem:
\begin{nth}
If we define 
\beq\label{mk5}
{\cal P}_n(x)=P_{n-1}(0)P_n(x)-P_n(0)P_{n-1}(x),\quad 
{\cal Q}_n(x)= P_{n-1}(0)Q_n(x)-P_n(0)Q_{n-1}(x), 
\eeq
then, for an indeterminate Stieltjes moment problem, we have
\beq\label{mk6}
\lim_{n\to\nf}\,\frac{{\cal Q}_n(x)}{{\cal P}_n(x)}=
\lim_{n\to\nf}\ \frac{\wh{F}_n(x)}{x\wti{F}_n(x)}=
\int\frac{d\psi_{0}(t)}{x-t},\qq x\in V={\mb C}\backslash{\mb R}\eeq
where the convergence is uniform for $x$ in any compact subset of $V.$
\end{nth}

\noindent{\bf Proof:}

\noindent This time we use two further relations given in \cite[p. 14]{Ak}:
\beq\label{mk7}\barr{l}
P_{n-1}(x)=Q_{n-1}(0)D_n(x)-P_{n-1}(0)B_n(x),\\[4mm]
Q_{n-1}(x)=Q_{n-1}(0)C_n(x)-P_{n-1}(0)A_n(x).\earr
\eeq
Combining (\ref{mk5}) and (\ref{mk6}) one gets
\beq\label{mk8}
{\cal Q}_n(x)=\rho_n\,C_n(x),\qq\qq {\cal P}_n(x)=\rho_n\,D_n(x),\qq 
\rho_n=\frac{(-1)^{n+1}}{\mu_n\sqrt{\pi_n}}.\eeq
In the limit $n\to\nf$ we have uniform convergence on compact subsets of $V$ to
\beq\label{mk9}
\lim_{n\to\nf}\,\frac{{\cal Q}_n(x)}{{\cal P}_n(x)}=\frac{C(x)}{D(x)},\qq x\in V.
\eeq
The theorem follows from relation (\ref{Ne3}). $\qq\Box$

We have given the proofs of Markov-like theorems in the modern setting due to 
Nevanlinna, however let us observe that in his own setting \cite{St2} Stieltjes 
was aware of the existence of the measures $\psi_0$ and $\psi_{\alf}$ and that 
they could be obtained from asymptotics. 

\section{A quartic example}
The polynomials $F_n(c,\mu;x)$ with recurrence coefficients
\beq\label{ex1}\barr{l}
\la_n=(4n+4c+1)(4n+4c+2)^2(4n+4c+3),\\[4mm]
 \mu_n=(4n+4c-1)(4n+4c)^2(4n+4c+1)+\mu\de_{n0},\earr
\qq c>0,\quad \mu\in{\mb R},\eeq
correspond to an indet S (hence indet H) moment problem. Their Nevanlinna matrix was 
given for $c=\mu=0$ in \cite{BV} and used to obtain the border measures $\psi_0$ 
and $\psi_{\alf}$ in closed. In the general case the Nevanlinna matrix was given 
in \cite{Va2} but explicit measures are quite hard to get. We will show how one can  
recover the results for $c=\mu=0$ using the previous Markov-like theorems.

We first need some background material. Let us define the entire functions $\de_l(x)$ , 
sometimes called trigonometric functions of order 4
\beq\label{qe1}
\de_l(x)=\sum_{n=0}^{\nf}(-1)^n\frac{x^{4n+l}}{(4n+l)!},\qq l=0,1,2,3.
\eeq
Their derivatives are
\beq\label{qe2}
\de'_0=-\de_3,\quad \de'_1=\de_0,\quad \de'_2=\de_1,\quad \de'_3=\de_2
\quad\Rightarrow\quad \de^{(4)}+\de_l=0,\quad l=0,1,2,3.
\eeq
the last relation explains their name. We have two simple cases
\beq\label{qe3}
\de_0(x)= \cos\left(\frac x{\sqrt{2}}\right)\cosh\left(\frac x{\sqrt{2}}\right),\qq
\de_2(x)= \sin\left(\frac x{\sqrt{2}}\right)\sinh\left(\frac x{\sqrt{2}}\right).
\eeq
We will need also the conformal mapping
\beq\label{qe4}
\tht(w)=\int_0^w\frac{du}{\sqrt{1-u^4}},\qq \tht(1)=K_0,\eeq
which maps $\ {\mb C}\backslash \cup_{k=0}^3\,i^k[1,\nf[\ $ onto the square with 
corners $\ \dst\pm \frac{K_0}{\sqrt{2}}\pm i\frac{K_0}{\sqrt{2}}.$
The inversion of the mapping $\tht(w)$ involves lemniscate elliptic functions, 
i. e. with parameter $k^2=1/2$ see \cite[p. 524]{WW} according to
\beq\label{qe5}
w(\tht)=\frac 1{\sqrt{2}}\frac{\sn\,(\sqrt{2}\tht)}{\dn\,(\sqrt{2}\tht}.\eeq

The basic tool will be the generating function
\beq\label{qe6}
\sum_{n\geq 0}\frac{(c+1)_n}{(c+1/2)_n}\frac{w^{4n+4c+1}}{(4c+1)!}F_n(c,\mu;x)=
{\cal F}(c,\mu;x;w),\eeq
with
\beq\label{qe7}\barr{l}\dst 
{\cal F}(c,\mu;x;w)=
\int_0^w \frac{\de_1(\rho(\tht(w)-\tht(u)))}{\rho}\frac{u^{4c-1}}{(4c-1)!}\,d\tht(u)
\\[5mm]\dst 
\hspace{4cm}+\mu_0\int_0^w\frac{\de_3(\rho(\tht(w)-\tht(u)))}{\rho^3}
\frac{u^{4c+1}}{(4c+1)!}\,d\tht(u),\earr\qq\rho=x^{1/4}.\eeq 
Asymptotic analysis gives
\beq\label{qe8}
F_n(c,\mu;x)\sim\frac{(4c+1)!}{4n+4c+1}\frac{(1/2)_n(c+1/2)_n}{n!\,(c+1)_n}\,
{\cal G}(c,\mu;x),\eeq
with
\beq\label{qe9}\barr{l}\dst 
{\cal G}(c,\mu;x)=\int_0^1 \de_0(\rho(\tht(1)-\tht(u)))
\frac{u^{4c-1}}{(4c-1)!}\,d\tht(u)\\[5mm]\dst 
\hspace{4cm}+\mu_0\int_0^1\frac{\de_2(\rho(\tht(1)-\tht(u)))}{\rho^2}
\frac{u^{4c+1}}{(4c+1)!}\,d\tht(u).\earr\eeq 
So, denoting by $F_n(x)$ the polynomials corresponding to the case $c=\mu=0$ we get, 
by a limiting process
\beq\label{qeas1}
F_n(x)\sim \pi_n(c=0)\ \de_0\left(x^{1/4}K_0/\sqrt{2}\right).\eeq
The asymptotics of $F_n^{(1)}(x)=F_n(c=1,\mu=0;x)$ is also easily obtained
\beq\label{qe11}
\frac{F_{n-1}^{(1)}(x)}{\mu_1}\sim \pi_n(c=0)\int_0^1 
\frac{\de_2(x^{1/4}(\tht(1)-\tht(u)))}{x^{1/2}}\,u\,d\tht(u).\eeq
Going first to the variable $\tht$ and then to $\sqrt{2}(\tht(1)-\tht)$ we 
are left with
\beq\label{qeas2}
\frac{F_{n-1}^{(1)}(x)}{\mu_1}\sim -\pi_n(c=0) \int_0^{K_0} 
\frac{\de_2(x^{1/4}u/\sqrt{2}))}{x^{1/2}}\,\cn u\ \frac{du}{\sqrt{2}}.\eeq
So we can state, for the Friedrichs extension of the Jacobi matrix:

\begin{nth}
The Stieltjes transform of the measure   
for $\,F_n(x)\equiv F_n(c=0,\mu=0;x)\,$ reads
\beq
\int\frac{d\psi_{\alf}(t)}{x-t}=\frac 1{\de_0(x^{1/4}u/\sqrt{2})}\int_0^{K_0} 
\frac{\de_2(x^{1/4}u/\sqrt{2}))}{x^{1/2}}\,\cn u\ \frac{du}{\sqrt{2}},\eeq
and the measure 
\beq\label{fr}
\psi_{\alf}=\frac{4\pi}{K_0^2}\sum_{n=0}^{\nf}\frac{(2n+1)\pi}{\sinh((2n+1)\pi)}\,
\eps_{x_n},\qq x_n=\left(\frac{(2n+1)\pi}{K_0}\right)^4.
\eeq
\end{nth}

\noindent{\bf Proof:}

\noindent The Stieltjes transform follows from (\ref{qeas1}), (\ref{qeas2}) and 
the first Markov-like theorem. The jumps occur at
\beq
x_n=\left(\frac{(2n+1)\pi}{K_0}\right)^4,\qq n\in{\mb Z}.
\eeq
To compute the masses one has to use the relation  proved in \cite[appendix]{Va1}
\beq
\int_0^{K_0}\de_2(x_n^{1/4}u/\sqrt{2})\,\cn u\,du=\frac 14 \int_{-K_0}^{+K_0}
\cos\left(x_n^{1/4}u/2\right)\frac{\cn u}{\dn u}\,du,
\eeq
and this last integral is easily computed from the Fourier series of the elliptic 
functions. It restricts $n$ to be positive, and gives 
\beq
\psi_n=\frac{4\pi}{K_0^2}\,\frac{(2n+1)\pi}{\sinh((2n+1)\pi)}
\qq n\geq 0,
\eeq
which ends the proof. $\qq\Box.$

Let us consider now the dual polynomials  
$\ \wti{F}_n(x)=F_n(c=1/2,\mu=12;x).$ Relation (\ref{qe8}) gives
\beq\label{qeas3}
\wti{F}_n(x)\sim 3\pi_n(c=0)\frac{\de_2(x^{1/4}K_0/\sqrt{2})}{x^{1/2}}.
\eeq
Similarly we have $\ \wh{F}_n=F_n(c=1/2,\mu=0;x)\ $ with the asymptotics
\beq\label{qeas4}
\wh{F}_n(x)\sim -3\pi_n(c=0)\int_0^{K_0}\de_0(x^{1/4}u/\sqrt{2})\,
\cn u\,\frac{du}{\sqrt{2}}.
\eeq
So we can state, for Krein's extension of the Jacobi matrix:

\begin{nth}
The Stieltjes transform of the measure for $\,F_n(x)\equiv F_n(c=0,\mu=0;x)\,$ reads
\beq
\int\frac{d\psi_0(t)}{x-t}=\frac 1{x^{1/2}\de_2(x^{1/4}u/\sqrt{2})}\int_0^{K_0} 
\frac{\de_0(x^{1/4}u/\sqrt{2}))}{x^{1/2}}\,\cn u\ \frac{du}{\sqrt{2}},\eeq
and the measure 
\beq\label{kr}
\psi_0=\frac{\pi}{K_0^2}\,\eps_{x_0}+\frac{4\pi}{K_0^2}\sum_{n=1}^{\nf}
\frac{2n\pi}{\sinh(2n\pi)}\,\eps_{x_n},\qq x_n=\left(\frac{2n\pi}{K_0}\right)^4.
\eeq
\end{nth}

\noindent{\bf Proof:}

\noindent The Stieltjes transform follows from (\ref{qeas3}), (\ref{qeas4}) and 
the second Markov-like theorem. The jumps occur at
\beq
x_n=\left(\frac{2n\pi}{K_0}\right)^4,\qq n\in{\mb Z}.
\eeq
To compute the masses one has to use the relation  proved in \cite[appendix]{Va1}
\beq
\int_0^{K_0}\de_0(x_n^{1/4}u/\sqrt{u})\,\cn u\,du=\frac 14 \int_{-K_0}^{+K_0}
\cos\left(x_n^{1/4}u/2\right)\frac 1{\dn u}\,du,
\eeq
and this last integral is easily computed from the Fourier series of the elliptic 
functions. It restricts $n$ to be positive, and gives 
\beq
\psi_n=\frac{4\pi}{K_0^2}\frac{2n\pi}{\sinh(2n\pi)},\qq n\geq 0,
\eeq
which ends the proof. $\qq\Box.$

\end{document}